\documentclass{iauc}
\usepackage{graphicx}
\pubyear{2005}
\volume{199}
\pagerange{1--}
\jname{Probing Galaxies through Quasar Absorption Lines}
\editors{P. R. Williams, C. Shu, and B. M\'{e}nard, eds.}

\newcommand{\ggkHI}{\hbox{{\rm H}\kern 0.1em{\sc i}}}
\newcommand{\ggkHII}{\hbox{{\rm H}\kern 0.1em{\sc ii}}}
\newcommand{\ggkLya}{\hbox{{\rm Ly}\kern 0.1em$\alpha$}}
\newcommand{\ggkLyb}{\hbox{{\rm Ly}\kern 0.1em$\beta$}}
\newcommand{\ggkMgI}{\hbox{{\rm Mg}\kern 0.1em{\sc i}}}
\newcommand{\ggkMgII}{\hbox{{\rm Mg}\kern 0.1em{\sc ii}}}
\newcommand{\ggkFeII}{\hbox{{\rm Fe}\kern 0.1em{\sc ii}}}
\newcommand{\ggkCII}{\hbox{{\rm C}\kern 0.1em{\sc ii}}}
\newcommand{\ggkCIII}{\hbox{{\rm C}\kern 0.1em{\sc iii}}}
\newcommand{\ggkCIV}{\hbox{{\rm C}\kern 0.1em{\sc iv}}}
\newcommand{\ggkNV}{\hbox{{\rm N}\kern 0.1em{\sc v}}}
\newcommand{\ggkOVI}{\hbox{{\rm O}\kern 0.1em{\sc vi}}}
\newcommand{\ggkSiIV}{{\rm Si}\kern 0.1em{\sc iv}}
\newcommand{\ggkSiII}{\hbox{{\rm Si}\kern 0.1em{\sc ii}}}
\newcommand{\ggkNII}{\hbox{{\rm N}\kern 0.1em{\sc ii}}}
\newcommand{\ggkHa}{\hbox{{\rm H}\kern 0.1em$\alpha$}}

\title[{\ggkMgII} Absorbing Galaxy Halos] 
{Galaxy Morphology -- Halo Gas Connections}

\author[Kacprzak, Churchill, \& Steidel]   
{Glenn G. Kacprzak$^1$,
Christopher W. Churchill$^1$ 
\break \and Charles C. Steidel$^2$}

\affiliation{$^1$Department of Astronomy, New Mexico State University, \break 
Las Cruces, NM 88003, USA \break 
email: glennk@nmsu.edu,cwc@nmsu.edu \\[\affilskip]
$^2$Department of Astronomy, California Institute of Technology, \break 
Pasadena, CA 91125, USA \break 
email: ccs@astro.caltech.edu}

\begin{document}

\maketitle

\begin{abstract}
We studied a sample of 38 intermediate redshift {\ggkMgII}
absorption--selected galaxies using (1) Keck/HIRES and VLT/UVES quasar
spectra to measure the halo gas kinematics from {\ggkMgII} absorption
profiles and (2) {\it HST}/WFPC--2 images to study the absorbing
galaxy morphologies. We have searched for correlations between
quantified gas absorption properties, and host galaxy impact
parameters, inclinations, position angles, and quantified
morphological parameters. We report a $3.2~\sigma$ correlation between
asymmetric perturbations in the host galaxy morphology and the
{\ggkMgII} absorption equivalent width.  We suggest that this
correlation may indicate a connection between past merging and/or
interaction events in {\ggkMgII} absorption--selected galaxies and the
velocity dispersion and quantity of gas surrounding these galaxies.


\end{abstract}

\firstsection 

\section{Introduction}

Our knowledge of the galaxy--halo environment is slowing being painted
by the study of the interstellar medium in both emission and
absorption. Understanding the distribution and kinematics of
extra--planer and extended halo gas, compared to that of the host
galaxy, can provide constraints on formation and evolutionary models.

Diffuse ionized gas (DIGs), usually observed in {\ggkNII} and {\ggkHa}
emission, has been studied in edge--on disk
galaxies. \cite{ggkref:lynds63} found filaments in M82 that flow away
from the plane of the disk. \cite{ggkref:rand00} showed that DIGs
extend out to 13 kpc above the plane of NGC 5775. The gas is
decreasing in rotational velocity with height above the plane and may
have no rotation above several kpcs.

In 21--cm emission, \cite{ggkref:swaters97} finds gas extending out to
at least 5 kpc above the galaxy plane of NGC 891 that rotates 25 to
100 km s$^{-1}$ more slowly than the disk gas.  Models are consistent
with a halo that is lagging behind the disk rotation. Fraternali
(\cite{ggkref:fraternali01}, 2002, 2004) also found halo gas with
slower rotation than the disk gas in several galaxies. This so--called
``anomalous gas'' displays a radial inward flow toward the center of
the galaxy. They may be observing the infalling stage of galactic
fountains.  They also detect ``forbidden gas'' that moves contrary to
disk rotation and does not fit well in the classical fountain
picture. Vertical motions and ``holes'' in the {\ggkHI} distribution
have been seen in M31 (\cite{ggkref:brinks86}) and in the dwarf
irregular HoII (\cite{ggkref:puche92}). These holes might be produced
by the expansion of large bubbles around stellar associations via
strong outflowing winds and/or supernovae. It is clear that the
disk--halo interface is dynamically and kinematically complex.
 
The study of halo gas in emission has several observational
challenges. Halo gas is intrinsically faint which limits studies to
the local universe and the present epoch. Furthermore, only the
highest column density regions of inner halos can be studied. An
additional complication is the distinction between halo and disk
emission for various galaxy orientations. On the other hand, observing
halo gas in absorption allows one to probe both the inner and outer
halos to much lower column densities at all redshifts. Absorption
selected galaxies are chosen by gas cross section only and not by {\it
a priori\/} knowledge of galaxy morphology, orientation, and surface
brightness.

Absorption studies primarily use the {\ggkMgII} $\lambda\lambda 2796,
2803$ doublet as a tracer of halo gas. Since the early 1990s, the
association of {\ggkMgII} absorption with normal, bright, field
galaxies has been well established (e.g., Bergeron \& Boiss{\' e}
1991; \cite{ggkref:steidel94}).

\begin{figure}
\includegraphics[width=5.4in,angle=0]{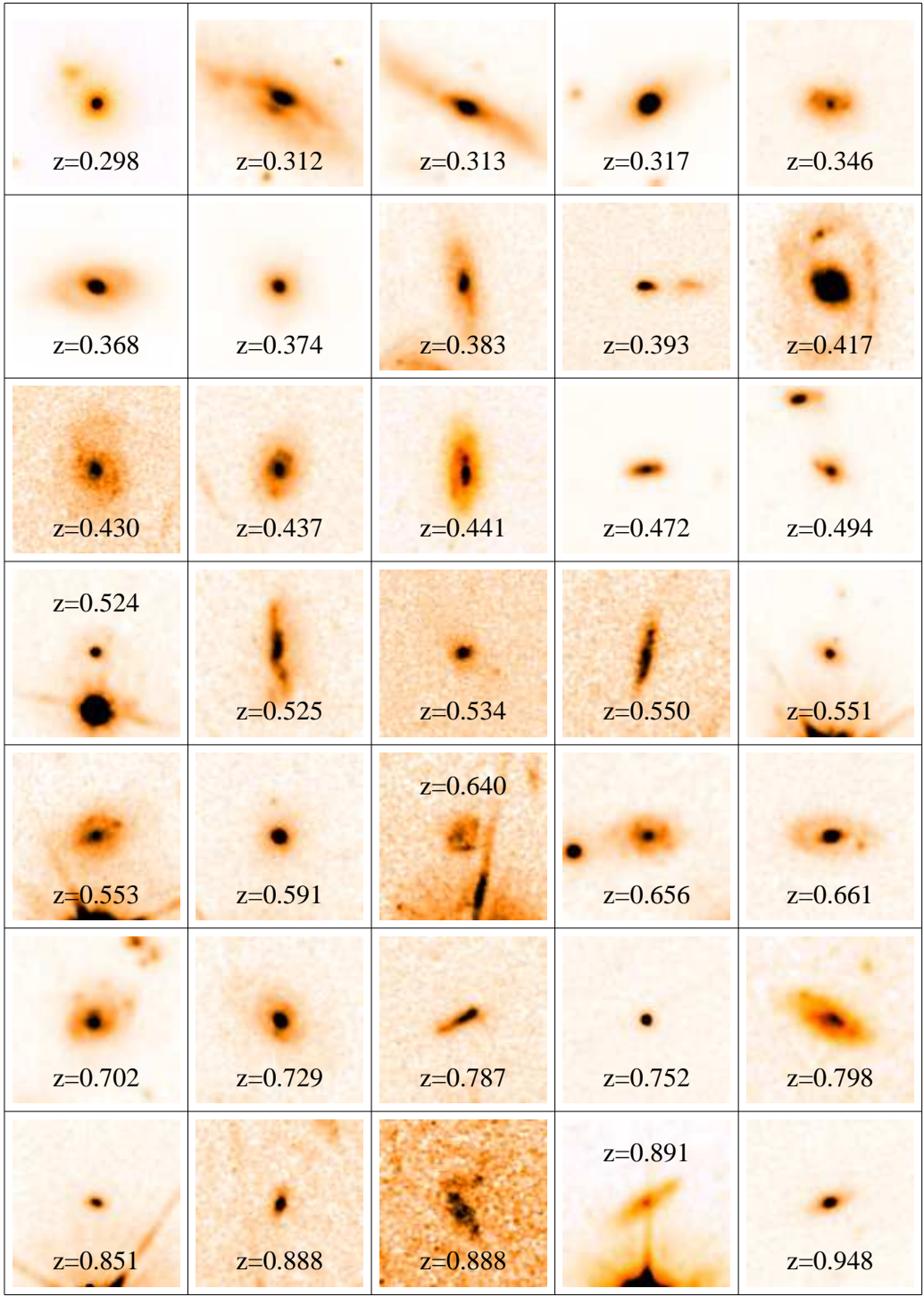}
\caption{{\it HST}/WFPC--2 images of 36 spectroscopically confirmed
{\ggkMgII} absorbing galaxies.  The ``postage stamps'' are 5$'' \times
5''$ and are orientated such that the quasar is down. Displaying the
galaxies in this fashion allows one to see the variety of galaxy
orientations with respect to the quasar line of sight.}
\label{ggkfig:mosaic}
\end{figure}

\begin{figure}
\includegraphics[width=5.3in,angle=0]{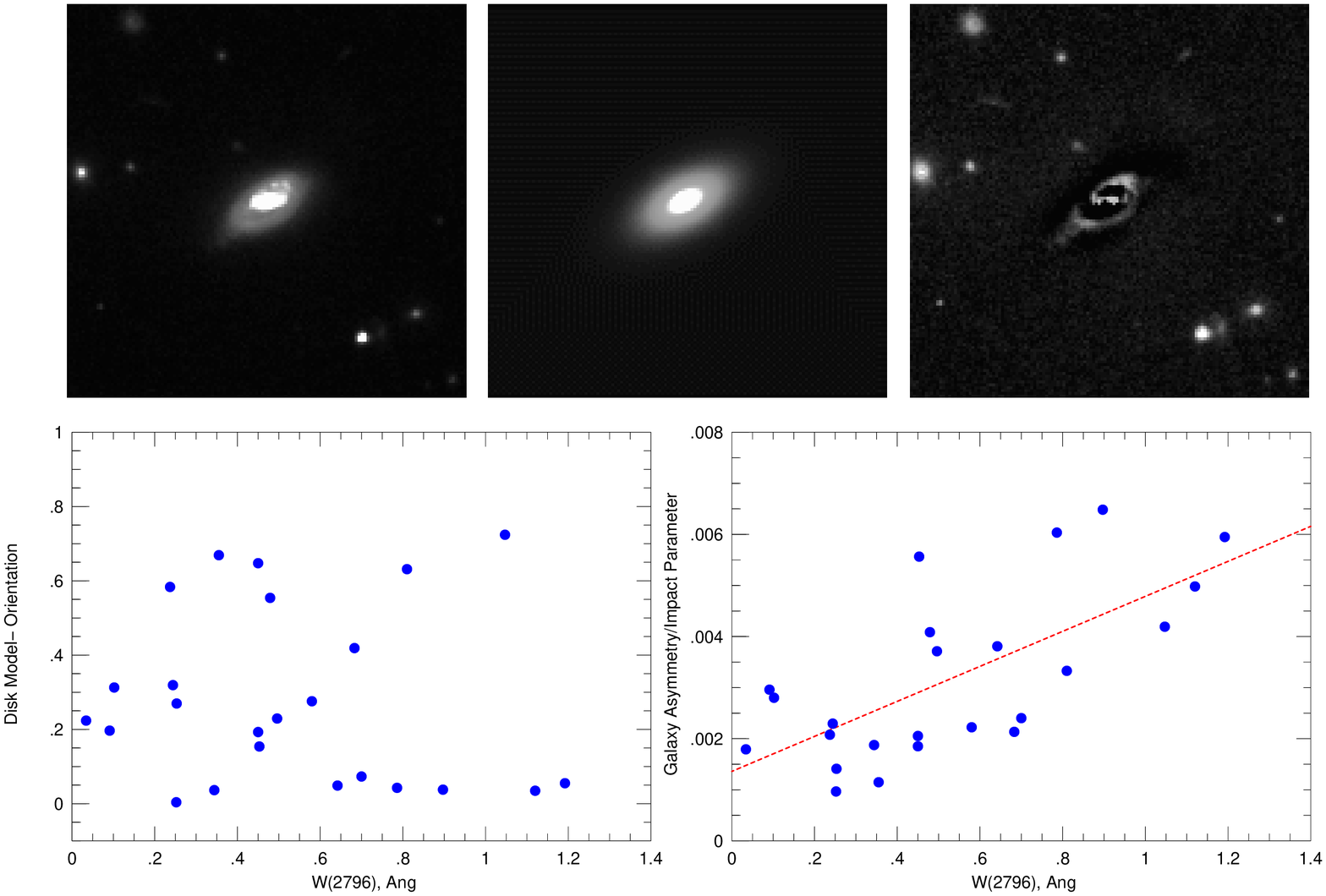}
\caption{ --- (top--left) {\it HST\/} image of an absorbing galaxy.
  --- (top--middle) Smooth galaxy model produced with GIM2D.  ---
  (top--right) Model residual image.  --- (bottom--left) Disk model
  galaxy orientation plotted against rest--frame {\ggkMgII} $\lambda
  2796$ equivalent width.  --- (bottom--right) Asymmetries in galaxy
  morphologies normalized to impact parameter plotted against
  {\ggkMgII} $\lambda 2796$ equivalent width.}
\label{ggkfig:asym}
\end{figure}

\section{Galaxy Orientations and Morphologies}

We have examined the detailed connections between halo gas and
galaxies using HIRES and UVES quasar spectra to study the {\ggkMgII}
absorption kinematics and {\it HST}/WFPC--2 images of the quasar
fields to measure the host galaxy properties. All galaxies in our
sample are spectroscopically confirmed to have the same redshift as
the {\ggkMgII} absorption.

In Figure~\ref{ggkfig:mosaic}, we present 36 of 38 galaxies in our
current sample having redshifts between 0.3 $< z <$ 1. Each galaxy
``postage stamp'' is $5^{\prime\prime} {\sf x}~5^{\prime\prime}$ with
the quasar oriented downward. The range of impact parameters are
$7\leq D \leq 80~h^{-1}$ kpc. Note that the galaxies exhibit a wide
range of orientations with respect to the quasar line of sight. Also,
there are a wide variety of galaxy morphologies. Some of the galaxies
appear slightly perturbed and/or have bright {\ggkHII} regions. Others
have minor satellites or major companions. Three of the {\ggkMgII}
absorbers are associated with double galaxies that could be in the
process merging or being harassed.

We modeled the morphology and orientations of the galaxies using GIM2D
(\cite[Simard {\etal} 2002]{ggkref:simard02}). The models comprise a
smooth disk and bulge component from which we quantify disk
inclinations, disk position angles, disk scale lengths, bulge position
angles, bulge effective radii, bulge--to--total ratios, galaxy
half--light radii, and morphological asymmetries. An example model and
the model residual are shown in the upper panels of
Figure~\ref{ggkfig:asym}. We used Voigt profile fitting to model the
HIRES and VLT spectra (\cite{ggkref:cv01}; \cite{ggkref:cvc03}). The
fits to the spectra provide the number of clouds in an absorption
system and their associated column densities, Doppler parameters, and
velocities. Directly from the spectra we also measured the equivalent
width, velocity spread, and velocity asymmetry of the absorbing gas.

To examine halo geometry, we tested for correlations between galaxy
orientation and absorption properties for a subsample having a
rest--frame equivalent width, $W(2796)$, less than 1.4 \AA. This
cutoff limits our subsample to ``classical'' systems and is designed
to eliminate DLAs and wind driven systems (see \cite{ggkref:cwc00};
\cite{ggkref:bond01b}). The orientation of a galaxy is the combined
projection of the galaxy's inclination, $i$, and the position angle,
$\phi$.  The position angle is defined as the primary angle between
the galaxy major axis and the quasar line of sight.  We find there are
no statistically significant correlations between galaxy orientation
parameters and {\ggkMgII} absorption properties.  In particular, if
the gas is distributed in a disk geometry then $\cos \phi \cos i$
should correlate with $W(2796)$. As shown in the bottom--left panel of
Figure~\ref{ggkfig:asym}, the distribution is consistent with being
random.

The GIM2D model residual image of the example galaxy, shown at the
top--right panel of Figure~\ref{ggkfig:asym}, displays a spiral barred
structure along with an extend tidal tail on one side. Without
modeling the galaxy these underlying features would go unnoticed. To
quantify these asymmetries we used two different methods. One method
computes the asymmetries on the residual image
(\cite{ggkref:schade95}) and the other is computed directly for the
science image (Abraham {\etal} 1994; 1996).  We find a $3.2~\sigma$
correlation using the Abraham {\etal} method and a $2.7~\sigma$
correlation using the \cite[Schade {\etal}]{ggkref:schade95} method
between $W(2796)$ and galaxy morphological asymmetry normalized by the
impact parameter. The correlation is shown in the bottom--right panel
of Figure~\ref{ggkfig:asym}, where the dashed line is a maximum
likelihood fit. In the near future, we will be conducting a
multivariate analysis to further explore the nature of correlations
between galaxy and absorption properties (e.g., \cite{ggkref:cwc00}).

\begin{figure}
\includegraphics[width=5.3in,angle=0]{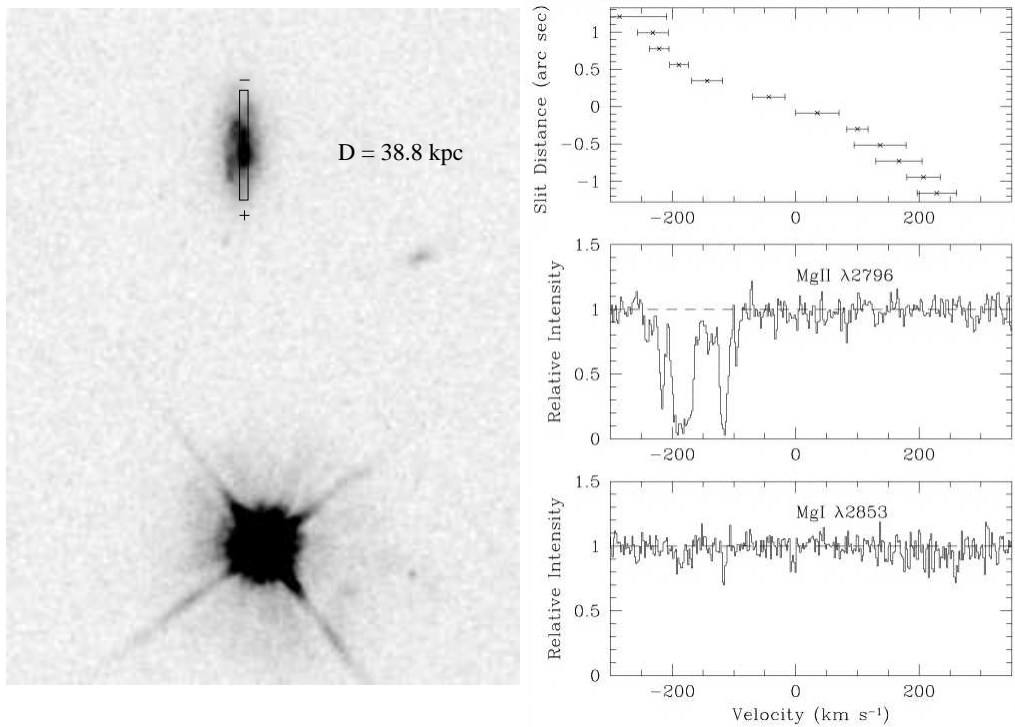}
\caption{--- (left) {\it HST}/WFPC--2 F702W images of Q1038+064. The
slit position used for the galaxy spectroscopy is indicated; the
relative spatial position along the slit is indicated with ''$+$'' and
''$-$''. The projected distance $D$ is 38.8 $h^{-1}$ kpc. --- (right)
Galaxy rotation curve (top) as a function of slit position with the
{\ggkMgII} kinematics of the observed {\ggkMgII} $\lambda 2796$
(middle) and {\ggkMgI} $\lambda 2853$ absorption profiles (bottom) from
the HIRES spectrum of the quasar.}
\label{ggkfig:kinematics}
\end{figure}

\section{Interpretations}

The lack of correlations with galaxy orientation implies that gaseous
halos are not necessarily disk--like or spherical or that the
distribution of gas could be patchy and have a less than unity
covering factor (also see Churchill {\etal} 2005 {\it this
volume}). The 3.2 $\sigma$ correlation suggests that a galaxy with
strong perturbations could produce a similar absorption strength at a
large galactocentric distance as a galaxy with mild perturbations at a
smaller galactocentric distance. It is possible that strong star
formation, winds, or tidal stripping could lead to the development of
both morphological perturbations in the galaxy and an increase in the
halo gas cross section. Galaxies that are more symmetric may not
produce as significant {\ggkMgII} absorption since there is a lacking
or inadequate mechanism for repopulating the halo gas. This would
explain why some bright galaxies close to the quasar produce very weak
to no absorption (see \cite{ggkref:cks05}). 

To help understand these results and to see if there may be sample
selection effects, we will employ more advanced models (i.e., lagging
halos, fountains, infall).  Furthermore, a full census of absorbers and
non--absorbers is central to a deeper interpretation of our results.

\section{Future of Kinematic Studies}

In order to further analyze the galaxy--halo kinematic connection it is
necessary that we obtain spectra of the galaxies. We can then place
the halo absorption features in the same velocity frame of the
galaxy. The velocity field between the galaxy and the halo will tell
us how they are kinematically related so that we can differentiate
between competing halo models.

A prime example of such a study was by \cite{ggkref:steidel02}, who
presented the kinematic properties of halo gas for a sample of five
edge--on galaxies. In Figure~\ref{ggkfig:kinematics} (left panel) we
present the {\it HST\/} image of the Q1038+064 field with an absorber
at $z =$0.4415. Long slit spectroscopy was obtained to get the
kinematics of the gas in the disk (upper--right panel). A HIRES
spectrum of the quasar was obtained for kinematic information of the
halo gas (lower--right panels). The halo gas kinematics is probed at
38.8 $h^{-1}$ kpc away from the disk. Absorption between $-$250 and
$-$150 km~s$^{-1}$ shows halo gas kinematics similar to that in the
disk.  Additional absorption between $-150$ and $-75$ km~s$^{-1}$ is
suggestive of slower rotation with respect to the disk, similar to
that mentioned by Fraternali {\etal} (2001, 2002, 2004). Thus, in this
galaxy probing out to distances far beyond the plane, one finds that
the halo gas is aware of the the kinematics of the disk.

In four of their five cases, \cite{ggkref:steidel02} found disk--like
halo kinematics consistent with a rotating and/or lagging halo.
Another study of and edge--on galaxy (\cite{ggkref:ellison03}) found
{\ggkMgII} absorption with velocities inconsistent with the galaxy
rotation. This ``forbidden'' gas may actually be caused by
superbubbles (see \cite{ggkref:bond01a}).

Over all, only a half dozen galaxies have been studied in great detail
by comparing the kinematics of the absorbing gas and the galaxy
kinematics.  However, these galaxies were all edge--on disk galaxies.
We are carrying out an in depth study, similar to that mentioned
above, for our sample of 38 galaxies having a wide range of
morphologies and line--of--sight orientations. Here, we have reported
an orientation and morphology analysis, in absence of galaxy spectra.
Our results hint that there are some relationships between the halo
and the host galaxy when looking at the distance normalized galaxy
perturbations. Obtaining the galaxy spectra will allow us to perform a
full halo model analysis which is much need in order to understand the
nature of galactic halos.

\begin{acknowledgments}

We would like to acknowledge a small grant from the IAU and Sigma Xi
Scientific Research Society. Michael Murphy, Michael Rauch, Wal
Sargent and Alice Shapley all made important contributions to the new
work presented herein.

\end{acknowledgments}

\end{document}